\documentclass[journal,comsoc]{IEEEtran}
\def\BibTeX{{\rm B\kern-.05em{\sc i\kern-.025em b}\kern-.08em
		T\kern-.1667em\lower.7ex\hbox{E}\kern-.125emX}}
\usepackage{float}
\normalsize

\usepackage{cite}
\usepackage{graphicx}
\ifCLASSINFOpdf
\else
\fi

\usepackage[cmex10]{amsmath}
\usepackage{relsize}
\usepackage{mathtools}

\DeclarePairedDelimiter{\norm}{\lVert}{\rVert}%
\usepackage{amsthm,amssymb}

\allowdisplaybreaks
\usepackage{longtable}
\usepackage{algorithmic}
\usepackage{array}
\usepackage{mdwmath}
\usepackage{mdwtab,bm}
\usepackage{cases}
\usepackage{eqparbox}
\usepackage[tight,footnotesize]{subfigure}
\usepackage{stfloats}

\theoremstyle{remark}

\usepackage{algorithm} %format of the algorithm
\usepackage{algorithmic} %format of the algorithm

\usepackage{multirow}

\usepackage{xcolor}
\begin{document}
	\title{Windowed Decoding for Delayed Bit-Interleaved Coded Modulation}
	%	\author{Author 1, Author 2, Author 3, Author 4, Author 5}
	\author{ {Yihuan~Liao,
			Min~Qiu,
			and~Jinhong~Yuan}\\ \vspace*{-10mm}
		\thanks{The authors are with the School of Electrical Engineering and Telecommunications, University of New South Wales, Sydney, NSW 2052, Australia (e-mail: yihuan.liao@unsw.edu.au; min.qiu@unsw.edu.au; j.yuan@unsw.edu.au). The work was partially supported by the Australian Research Council (ARC) Discovery Projects under Grant DP190101363 and by the ARC Linkage Project under Grant LP170101196.}}% <-this % stops a space
	\maketitle
	%		\vspace{-20mm}
	\begin{abstract}
		Delayed bit-interleaved coded modulation (DBICM) generalizes bit-interleaved coded modulation (BICM) by modulating differently delayed sub-blocks of codewords onto the same signals. DBICM improves transmission reliability over BICM due to its capability of detecting undelayed sub-blocks with the extrinsic information of the decoded delayed sub-blocks. In this work, we propose a novel windowed decoding algorithm for DBICM, which uses the extrinsic information of both the decoded delayed and undelayed sub-blocks, to improve the detection on all sub-blocks. Numerical results show that the proposed windowed decoding significantly outperforms the original decoding.  
	\end{abstract}
	

	% \fi
	\IEEEpeerreviewmaketitle
		\vspace{-3mm}
	\section{Introduction}
	\IEEEPARstart{T}{o} meet the tremendous demand for data transmission and achieving high spectral efficiencies, coded modulation (CM), which combines high order modulation with channel coding \cite{lin2004error,ryan2009channel}, has become indispensable in modern communication systems. As a pragmatic approach in CM, bit-interleaved coded modulation (BICM) has been extensively investigated for many wireless and optical communication systems \cite{669123, 1566630, bicm_wireless, 4026714, 6844864, szczecinski2015bit}. Later, delayed BICM (DBICM) was introduced in \cite{7593082} to improve the transmission reliability over BICM. In DBICM, the sub-blocks of codewords from the previous time slots to the current time slots are modulated to the same signal sequence. As a result, the decoded delayed sub-blocks in the current time slot can be used to improve the detection of the undelayed sub-blocks in the succeeding time slots. Previous works have reported noticeable improvements of DBICM over BICM with low-density parity-check (LDPC) codes \cite{7593082, 8625331, 8689212, 8437452, 8989384, 9373632}, polar codes \cite{8580932}, and coded sparse code multiple access \cite{9091254}. 
	
	Recently, the design and analysis for DBICM delay schemes and LDPC codes were investigated in \cite{9373632}, where a coding gain about $0.7$ dB was observed for DBICM over BICM for uniform Gray labeled $64$-quadrature amplitude modulation (QAM). In \cite{8437452}, constellation labeling design for uniform QAM DBICM with iterative detection and decoding (DBICM-ID) was investigated. Specifically, DBICM-ID iteratively decodes and detects the received sequences at several consecutive time slots regardless of the detection and decoding order \cite{8437452}.
	
	We notice that both DBICM and spatially coupled (SC) codes \cite{5695130, 7152893, 8368318, 8357804, 9328182} have a similar coupling structure that the component codeword spreads over several consecutive time instances. Hence, reliable information propagates through the coupled chain to improve the decoding of each component codeword. However, the conventional decoding for DBICM only improves the detection of undelayed sub-blocks in a single shot \cite{7593082}. In addition, the detection of the delayed sub-blocks in DBICM is the same as that in BICM, i.e., without any a priori information. Inspired by the decoding of SC codes, we propose windowed decoding for DBICM to iteratively improve the detection of both the undelayed sub-blocks and the delayed sub-blocks by using the extrinsic information from decoding either the delayed sub-blocks or undelayed sub-blocks in detection. We also use a normal graph \cite{910573} to visualize the information propagation in the DBICM system. Numerical results demonstrate significant performance improvement of DBICM with windowed decoding over DBICM with its original decoding and BICM. Furthermore, the proposed decoding algorithm also have comparable performance to DBICM-ID while having a lower detection complexity. Simulation results show that using both existing and designed LDPC codes with different code rates and modulations, the proposed windowed decoding improves the performance over the conventional decoding for DBICM.
	
	Hereafter, we use normal case letters for constant and scalars, while vectors are represented by boldface letters. 
	\vspace{-3mm}
	\section{DBICM System Model} \label{sec:dbicm_system}
	\begin{figure}[h]
		%\normalsize
		\centering
		\vspace*{-3mm}
		\includegraphics[scale=0.48]{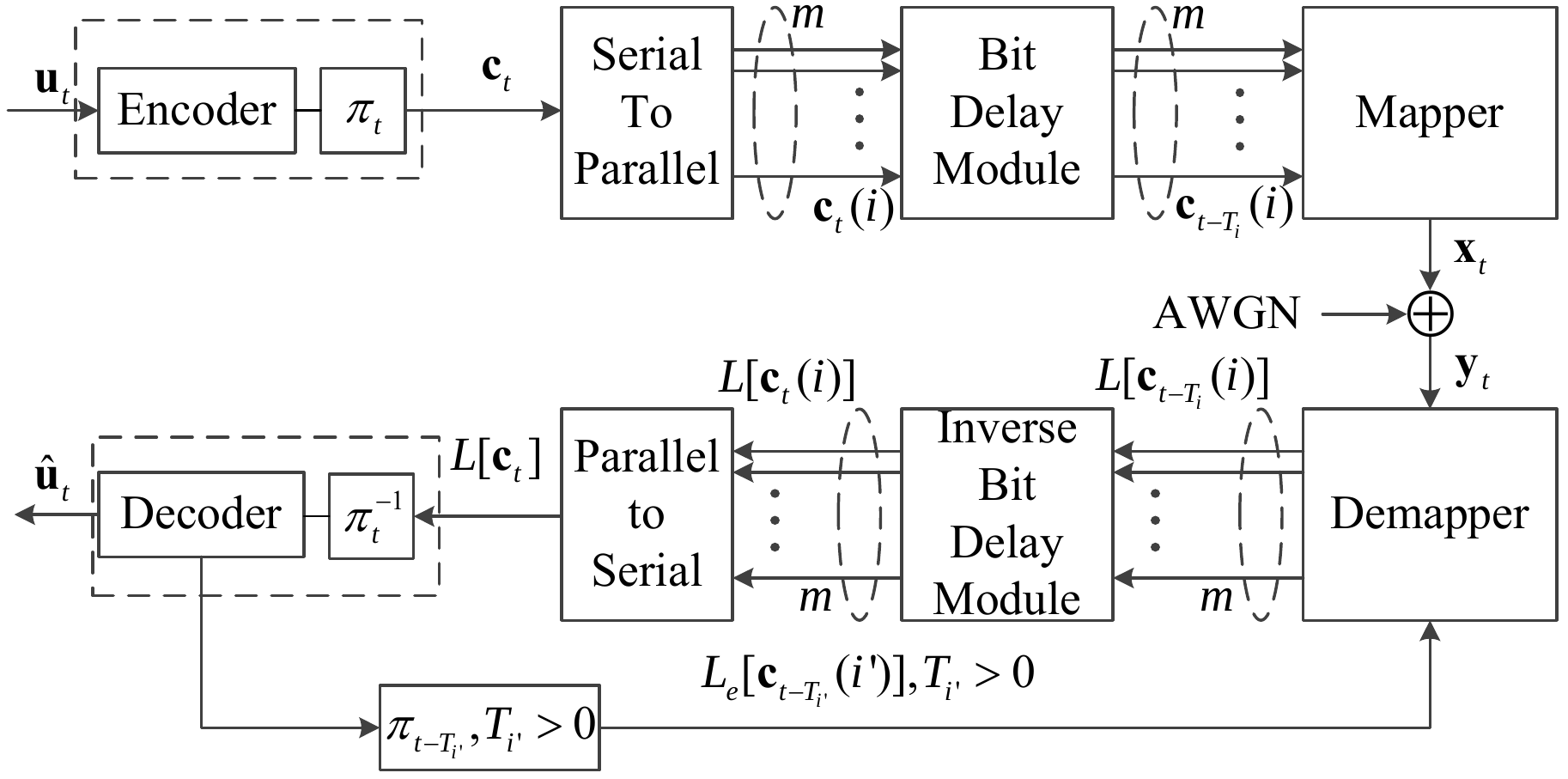}
						\vspace*{-5mm}
		\caption{Block diagram of DBICM structure.~~~~~~}
		\label{fig:DBICM_structure}
		\vspace*{-3mm}
	\end{figure}
	
	We consider a DBICM system with a $2^{m}$-ary complex signal constellation $\chi$ over an additive white Gaussian noise (AWGN) channel. The system model is depicted in Fig. \ref{fig:DBICM_structure}. At time $t$, an information sequence $\mathbf{u}_t$ of length $K$ is encoded and interleaved to become a codeword sequence $\mathbf{c}_t$ of length $N$. Following \cite{9373632}, the interleaver design is embedded into the LDPC code design. For simplicity, we assume that $N$ is divisible by $m$. Then, codeword $\mathbf{c}_t$ is equally divided into $m$ sub-blocks,  $\mathbf{c}_{t}(0), \cdots, \mathbf{c}_{t}(m-1)$, of length $n=N/m$. These $m$ sub-blocks are delayed by the bit delay module following the delay scheme $\mathbf{T}=[T_i]^{m-1}_{i=0}, T_{i}\in\{T_{\text{min}}, \cdots, T_{\text{max}}\}$, where $T_{\text{max}}$ and $T_{\text{min}}$ represent the maximum and minimum number of the delayed time slots, respectively. For simplicity, we further assume that $T_{\text{min}}=0$. The bit delay module then outputs $\mathbf{c}_{t-T_i}(i)$, $i\in\{0,1,\cdots,m-1\}$, which is the $i$-th sub-block of codeword $\mathbf{c}_{t-T_i}$ encoded at time $t-T_i$ and delalyed to time $t$. Next, the mapper modulates $m$ sub-blocks $\mathbf{c}_{t-T_0}(0), \cdots, \mathbf{c}_{t-T_{m-1}}(m-1)$ into a signal sequence $\mathbf{x}_t = [x_t^0, x_t^1,\cdots, x_t^{n-1}]$, where $x_t^j\in\chi$, $j\in\{0,\cdots, n-1\}$. From now on, for $i'\neq\bar{i'}$, and $i',\bar{i'}\in\{0, 1, \cdots, m-1\}$, we refer to $\mathbf{c}_{t-T_{i'}}(i')$ with $T_{i'}>0$ and $\mathbf{c}_{t-T_{\bar{i'}}}(\bar{i'})$ with $T_{\bar{i'}}=0$ (i.e., $c_t(\bar{i'}))$, as the delayed sub-block and the undelayed sub-block, respectively. The received signal sequence at time $t$ is $\mathbf{y}_t = \mathbf{x}_t + \mathbf{z}_t$, where $\mathbf{z}_t$ represents the AWGN noise samples with zero mean and variance $\sigma^2$ per real and imaginary dimension.  
	
	Let $L[\cdot]$ and $L_e[\cdot]$ denote the log-likelihood ratio (LLR) from demapping, and extrinsic information from decoding, respectively. At time $t$, for $i\in\{0,1,\cdots,m-1\}$ and $j\in\{0,1,\cdots,n-1\}$, we use $\hat{x}_{t}^j$ and $\hat{c}^{j}_{t-T_i}(i)$ to denote the estimation on the transmitted signal $x_t^j$ and $j$-th bit in $\mathbf{c}_{t-T_i}(i)$, respectively. The LLR of $c_{t-T_i}^j(i)$ given $y^j_t$ is 
	\begin{align} \label{eq:demo_without_ap}
		L[c_{t-T_i}^j(i)|y_t^j] = \ln\left(\dfrac{\sum\limits_{\{\hat{x}_t^j\in\chi|\hat{c}_{t-T_i}^j(i) = 0\}}e^{-\frac{\norm{{y}_t^j-{\hat{x}_t^j}}^2}{2\sigma^2}}}{\sum\limits_{\{\hat{x}_t^j\in\chi|\hat{c}_{t-T_i}^j(i) = 1\}}e^{-\frac{\norm{{y}_t^j-{\hat{x}_t^j}}^2}{2\sigma^2}}}\right).
	\end{align}
	At time $t$, DBICM enables the delayed sub-blocks $\mathbf{c}_{t-T_{i'}}(i')$ to be decoded prior to the undelayed sub-blocks $\mathbf{c}_{t-T_{\bar{i'}}}(\bar{i'})$. After decoding $\mathbf{c}_{t-T_{i'}}(i')$, their extrinsic information $L_e[\mathbf{c}_{t-T_{i'}}(i')]$ is passed to the demapper as the a priori information for the detection of $\mathbf{c}_{t-T_{\bar{i'}}}(\bar{i'})$. We denote the probability of $c^j_{t-T_i}(i)$ being $b\in\{0,1\}$ by $P_{b}(c^j_{t-T_{i}}({i}))$, where 
	\begin{align} \label{eq:prob}
		P_{b}(c^j_{t-T_{i}}(i)) = b+\frac{(-1)^b}{1+e^{-L_e[c^j_{t-T_{i}}(i)]}}.
	\end{align}  Given $L_e[c_{t-T_{i'}}^j({i'})]$, the LLR of the $j$-th coded bit in an undelayed sub-block $\mathbf{c}_{t}(\bar{i'})$ is updated following
	%	\vspace*{-2mm}
	\begin{align} \label{eq:demo_with_ap}
		&L[c_{t}^j(\bar{i'})|y_{t}^j,L_e[c_{t-T_{i'}}^j({i'})]] = \notag\\  &\ln\left(\dfrac{\sum\limits_{b=0}^{1}\sum\limits_{\{\hat{x}_{t}^j\in\chi|\hat{c}_{t}^j=0,\hat{c}_{t-T_{i'}}^j({i'}) = b\}}\hspace*{-10mm}e^{-\frac{\norm{{y}_{t}^j-{\hat{x}_{t}^j}}^2}{2\sigma^2}}P_{b}(c^j_{t-T_{i'}}({i'}))}{\sum\limits_{b=0}^{1}\sum\limits_{\{\hat{x}_{t}^j\in\chi|\hat{c}_{t}^j=1,\hat{c}_{t-T_{i'}}^j({i'})=b\}}\hspace*{-10mm}e^{-\frac{\norm{{y}_{t}^j-{\hat{x}_{t}^j}}^2}{2\sigma^2}}P_{b}(c^j_{t-T_{i'}}({i'}))}\right).
	\end{align} Finally, the decoder uses the LLR of the undelayed sub-blocks $L[c_{t}^j(\bar{i'})|y_{t}^j,L_e[c_{t-T_{i'}}^j({i'})]]$ from Eq. (\ref{eq:demo_with_ap}) and the LLR of the delayed sub-blocks $L[c_{t}^j(i')|y_{t+T_{i'}}^j]$ from Eq. (\ref{eq:demo_without_ap}) to estimate the transmitted information $\mathbf{u}_t$. Compared with BICM, DBICM additionally detects the undelayed sub-blocks once following Eq. (\ref{eq:demo_with_ap}), which improves the transmission reliability.
	
	%	While the delay allows DBICM to benefit from a performance improvement, it requires $T_{\max}$ extra time slots in each transmission frame as compared with BICM to transmit the same number of codewords and symbols, which provides a spectral efficiency loss.
	
	While the delay allows DBICM to benefit from a performance improvement, it also leads to a spectral efficiency loss. Consider a DBICM system that transmits $T_n$ time slots at a code rate $R=\frac{K}{N}$, the spectral efficiency is 
	\begin{align}\label{eq:spectral_eff_dbicm}
		\eta_{\text{DBICM}} = mR\left(\frac{T_n-T_{\text{max}}}{T_n}\right).
	\end{align} 
	To minimize the loss of spectral efficiency in DBICM, we consider $T_{\text{max}} = 1$ in this paper, which is sufficient to allow DBICM to approach the CM capacity, while outperforming BICM \cite{9373632}. Furthermore, to minimize the spectral efficiency loss in DBICM, we consider $T_n \gg T_{\text{max}}$ in this paper.
	
%	Consider a DBICM system that transmits $T_n$ time slots at a code rate $R=\frac{K}{N}$, the spectral efficiency is 
%	\begin{align}\label{eq:spectral_eff_dbicm}
%		\eta_{\text{DBICM}} = mR\left(\frac{T_n-T_{\text{max}}}{T_n}\right).
%	\end{align}
%	To minimize the loss of spectral efficiency in DBICM, we consider $T_{\text{max}} = 1$ in this paper, which is sufficient to allow DBICM to approach the CM capacity, while outperforming BICM \cite{9373632}. Furthermore, to minimize the spectral efficiency loss in DBICM, we consider $T_n \gg T_{\text{max}}$ in this paper.
	
	% comment this part in latex
	%	For example, at time $t+T_{\text{max}}$, the LLRs of all $m$ sub-blocks associated with the codeword sequence $\mathbf{c}_t$ are received and detected. Then, the inverse bit delay module groups the LLRs of these detected $m$ sub-blocks to become $L[\mathbf{c}_{t}(i)],i\in\{0,1,\cdots,m-1\}$. Then, the $m$-to-1 parallel to serial module transform $L[\mathbf{c}_{t}(i)],i\in\{0,1,\cdots,m-1\}$, into $L[\mathbf{c}_{t}]$, which is a sequence with length $N$, representing the LLRs of all coded bits in $\mathbf{c}_t$. Then, to recover $\mathbf{u}_t$, the decoder takes $L[\mathbf{c}_{t}]$ as the input to produce the estimation $\hat{\mathbf{u}}_t$. 
	
	Note that the original decoding of DBICM only updates the LLRs of the undelayed sub-blocks once via passing the extrinsic-information of the delayed sub-blocks forwardly to the undelayed sub-blocks \cite{7593082}. To improve the detection of all sub-blocks from BICM, we propose a windowed decoding algorithm for DBICM as described in the next section. 
%	\vspace*{-3mm}
	\section{Windowed Decoding for DBICM} \label{sec:win_dbicm}	
	
	In this section, we first look into the normal graph of a DBICM transmission scheme. We then introduce the windowed decoding algorithm for DBICM and compare the proposed windowed decoding with the original decoding for DBICM, as well as the decoding of DBICM-ID \cite{8437452}.
	\vspace*{-3mm}
	\subsection{Normal Graph} \label{sec:normal_graph}
\begin{figure}[h]
		%\normalsize
		\centering
		\vspace{-3mm}
		\includegraphics[width=0.48\textwidth]{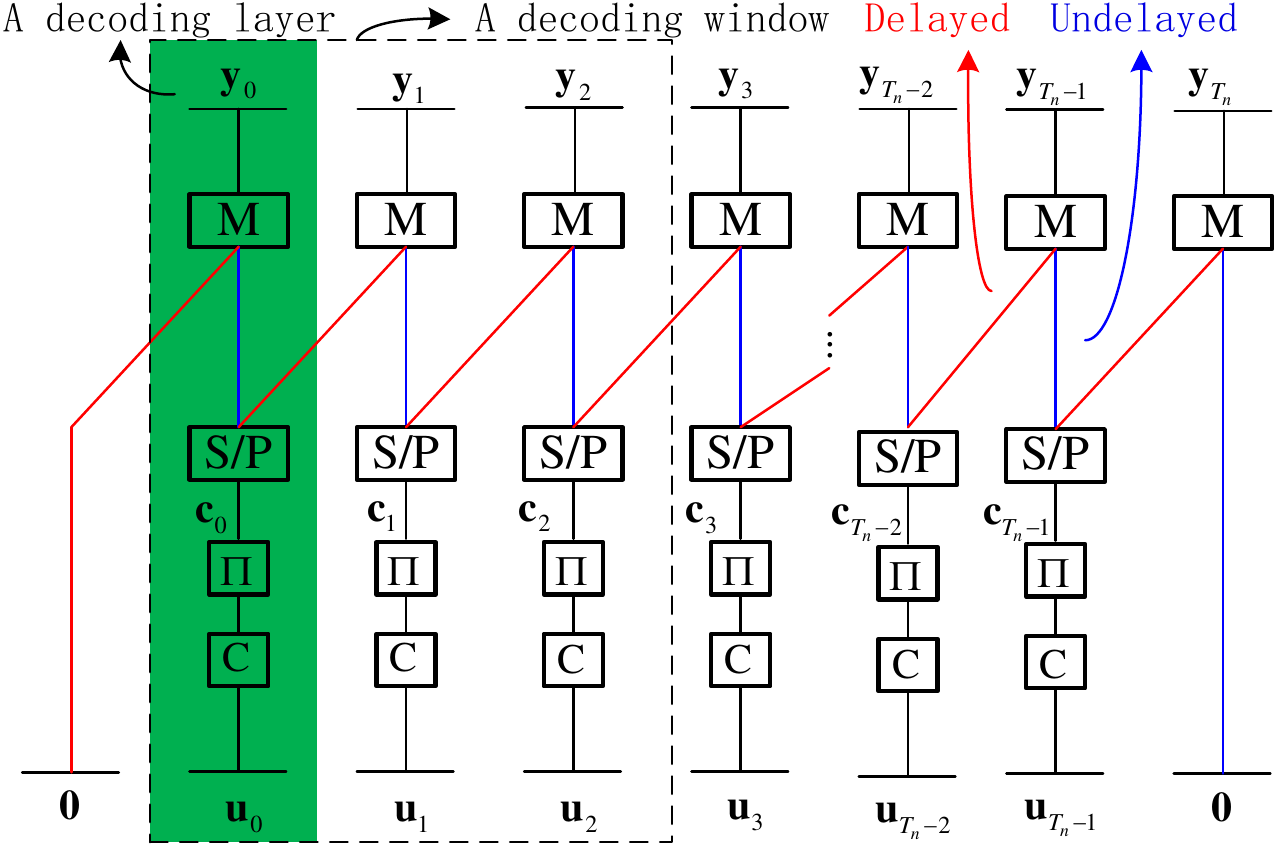}
		\caption{The normal graph of the DBICM scheme with $T_{\text{min}} = 0$, $T_{\text{max}} = 1$.~~~~~~}
		\vspace*{-3mm}
		\label{fig:DBICM_normal_graph}
	\end{figure}

	We show the message processing/passing of a DBICM transmission scheme with $T_{\text{min}} = 0$ and $T_{\text{max}} = 1$, via a normal graph \cite{910573} in Fig. \ref{fig:DBICM_normal_graph}. The message processors and variables are represented by square nodes and edges, respectively in the normal graph. For simplicity, we merge multiple variables of the same type into one edge. Furthermore, to ease the presentation, we use $\mathbf{u}_t$, $\mathbf{c}_t$, and $\mathbf{y}_t$ to represent their associated edges in the normal graph with slightly abuse the notation. In the following, we present the definitions of the four types of nodes in Fig. \ref{fig:DBICM_normal_graph}.
	
	\begin{enumerate}
		\item Node $\boxed{\text{C}}$ is defined as the constraint that $\mathbf{c}_t$ is encoded from $\mathbf{u}_t$. For decoding, node $\boxed{\text{C}}$ applies a decoding algorithm to decode $\mathbf{c}_t$ and compute its extrinsic information. 
		\item Node $\boxed{\Pi}$ is defined as the interleaver/deinterleaver which is self-explanatory.
		\item Node $\boxed{\text{S/P}}$ is defined as the serial-to-parallel/parallel-to-serial transformation which transforms the input messages between the serial and parallel form.
		\item Node $\boxed{\text{M}}$ is defined as the demapper which demodulates $\mathbf{y}_t$ into $m$ parallel sequences of LLRs. The demodulation on $\mathbf{y}_t$ and its adjacent signal sequences, $\mathbf{y}_{t+1}$ can be updated by using the extrinsic information associated with $\mathbf{c}_t$ as the a priori information. 
	\end{enumerate}	
	
\begin{figure}[h]
		\centering
		\vspace*{-3mm}
		\includegraphics[width=0.48\textwidth]{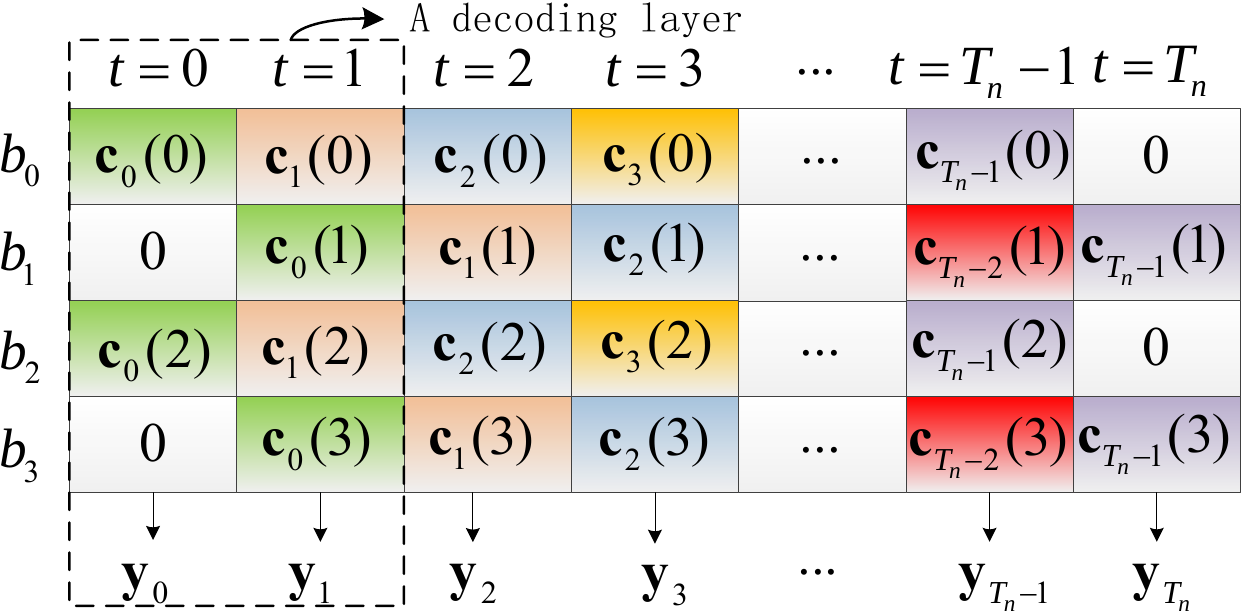}
		\vspace*{-3mm}
		\caption{An example of sub-blocks grouping for a $16$-QAM DBICM scheme with a delay scheme $\mathbf{T}=[0,1,0,1]$.~~~~~~}
		\vspace*{-3mm}
		\label{fig:delay_example}
	\end{figure}

	As shown in Fig. \ref{fig:DBICM_normal_graph}, a decoding layer of a DBICM system contains the nodes in one time slot and its connected edges. For $T_{\text{max}} = 1$, the connected edges in a decoding layer span three time slots in the normal graph. Furthermore, within a decoding layer, the signal sequences received in two consecutive time slots are required to be detected to decode a codeword. In our case, to recover $\mathbf{u}_t$, both $\mathbf{y}_t$ and $\mathbf{y}_{t+1}$ are required. Consider a DBICM using $T_n$ time slots. To initialize and terminate the DBICM transmission, known information is filled in sub-blocks $\mathbf{c}_{t-T_{i'}}(i')$ at time $t=0$, $T_{i'}>0$, and sub-blocks $\mathbf{c}_{t-T_{\bar{i'}}}(\bar{i'})$ at time $t=T_n$, $T_{\bar{i'}}=0$, respectively. For example, we show the sub-blocks grouping for a $16$-QAM DBICM with a delay scheme $\mathbf{T}=[0,1,0,1]$ in Fig. \ref{fig:delay_example}, where the known information are assumed to be all-zero. 
	
	For the original decoding of DBICM, the received signal sequence $\mathbf{y}_t$, the extrinsic information of $\mathbf{c}_{t-1-T_{i'}}(i')$, and the initial LLRs of $\mathbf{c}_{t-T_{\bar{i'}}}(\bar{i'})$ are input to the decoding layer at time $t$, which outputs codeword $\mathbf{c}_t$ and passes the extrinsic information of $\mathbf{c}_{t-T_{i'}}(i')$ to its next decoding layer at time $t+1$ via the delayed edge as shown in Fig. \ref{fig:DBICM_normal_graph}. In the example shown in Fig. \ref{fig:delay_example}, the extrinsic information of $\mathbf{c}_{1}(1)$ and $\mathbf{c}_{1}(3)$ as a result of decoding $\mathbf{c}_1$ at $t=2$ are used to improve the detection on $\mathbf{c}_{2}(0)$ and $\mathbf{c}_{2}(2)$. Then, $\mathbf{y}_{3}$ is detected without a priori information to produce the initial LLRs of $\mathbf{c}_{2}(1)$ and $\mathbf{c}_{2}(3)$. Finally, the estimation of codeword $\mathbf{c}_2$ is output from the decoder at time $t=3$. However, the extrinsic information of the delayed sub-blocks is only used once, and the extrinsic information of the undelayed sub-blocks is not exploited.
		\vspace*{-3mm}
	\subsection{Windowed Decoding for DBICM}
	
	\begin{table}[h]\label{table:algorithm 1} 
		\vspace*{-2mm}
		\begin{algorithm} [H]              
			\renewcommand\thealgorithm{1}
			\centering
			\caption{Windowed Decoding for DBICM}
			\label{algorithm:algorithm 1}    
			\begin{algorithmic} [1] 
				\REQUIRE Window size $W$, transmission frame $T_n$, maximum number of windowed iteration $I_{\text{max}}$, modulation level $m$, delay scheme $\mathbf{T}$, and received signal $\mathbf{y}_t, 0\le t\le T_n-1$.
				\ENSURE Estimated information $\hat{\mathbf{u}}_t, 0\le t\le T_n-1$.
				%					\STATE $L_e[\mathbf{c}_{-1}(i')]=\infty$ and $L_e[\mathbf{c}_{T_n}(\bar{i'})]=\infty$.
				\FOR {$t = 0, 1, \cdots, T_n-W$}
				\IF {$t=0$}
				\FOR {$w=1,\cdots,W-1$}
				\STATE Input $\mathbf{y}_{t+w}$ to $\boxed{\text{M}}$ and output $L^{(0)}[\mathbf{c}_{t+w-T_{0}}(0)|\mathbf{y}_{t+w}]$, $\cdots$, $L^{(0)}[\mathbf{c}_{t+w-T_{m-1}}(m-1)|\mathbf{y}_{t+w}]$ following Eq. (\ref{eq:demo_without_ap}). 
				\ENDFOR
				\ELSE 
				\STATE Input $\mathbf{y}_{t+W-1}$ to $\boxed{\text{M}}$ and output $L^{(0)}[\mathbf{c}_{t+W-1-T_{0}}(0)|\mathbf{y}_{t+W-1}]$, $\cdots$, $L^{(0)}[\mathbf{c}_{t+W-1-T_{m-1}}(m-1)|\mathbf{y}_{t+W-1}]$ following Eq. (\ref{eq:demo_without_ap}). 
				\ENDIF
				\FOR {$l = 1, \cdots, I_{\text{max}}$}
				\FOR {$w = 0, 1, \cdots, W-1$} % [$\rightarrow$ Forward recursion]
				\IF {$w>0$}
				\IF {$l = 1$}
				\STATE Input $\overrightarrow{L}^{(l)}[\mathbf{c}_{t+w-1}(\bar{i'})|\mathbf{y}_{t+w},L_e[\mathbf{c}_{t+w-2}(i')]]$ and $L^{(0)}[\mathbf{c}_{t+w-1}(i')|\mathbf{y}_{t+w}]$ to $\boxed{\text{C}}$ and output $L_e[\mathbf{c}_{t+w-1}]$.
				\ELSE
				\STATE Input $\overrightarrow{L}^{(l)}[\mathbf{c}_{t+w-1}(\bar{i'})|\mathbf{y}_{t+w},L_e[\mathbf{c}_{t+w-2}(i')]]$ and $\overleftarrow{L}^{(l-1)}[\mathbf{c}_{t+w-1}(i')|\mathbf{y}_{t+w},L_e[\mathbf{c}_{t+w}(\bar{i'})]]$ to $\boxed{\text{C}}$ and output $L_e[\mathbf{c}_{t+w-1}]$.
				\ENDIF 
				%					\STATE $\boxed{\text{C}}$ outputs $\hat{\mathbf{u}}_{t+w-1}$ and $L_e[\mathbf{c}_{t+w-1}]$.
				\ENDIF
				\IF {$t+w<T_n$}
				\STATE  Input $L_e[\mathbf{c}_{t+w-T_{i'}}(i')]$ to $\boxed{\text{M}}$ and output $\overrightarrow{L}^{(l)}[\mathbf{c}_{t+w-T_{\bar{i'}}}(\bar{i'})|\mathbf{y}_{t+w},L_e[\mathbf{c}_{t+w-T_{i'}}(i')]]$ following Eq. (\ref{eq:demo_with_ap}). 
				\ENDIF
				\ENDFOR
				\FOR {$w = W, W-1, \cdots, 1$} %[$\rightarrow$ Backward recursion]
				\STATE  Input $L_e[\mathbf{c}_{t+w-T_{\bar{i'}}}({\bar{i'}})]$ as a priori information to $\boxed{\text{M}}$ and produce $\overleftarrow{L}^{(l)}[\mathbf{c}_{t+w-T_{i'}}(i')|\mathbf{y}_{t+w},L_e[\mathbf{c}_{t+w-T_{\bar{i'}}}({\bar{i'}})]]$ following Eq. (\ref{eq:demo_with_ap_back}). 
				\STATE Input $\overrightarrow{L}^{(l)}[\mathbf{c}_{t+w-1}(\bar{i'})|\mathbf{y}_{t+w},L_e[\mathbf{c}_{t+w-2}(i')]]$ and $\overleftarrow{L}^{(l)}[\mathbf{c}_{t+w-1}(i')|\mathbf{y}_{t+w},L_e[\mathbf{c}_{t+w}(\bar{i'})]]$ to $\boxed{\text{C}}$ and output $L_e[\mathbf{c}_{t+w-1}]$.
				%					\STATE $\boxed{\text{C}}$ outputs $\hat{\mathbf{u}}_{t+w-1}$ and $L_e[\mathbf{c}_{t+w-1}]$.
				\ENDFOR
				\IF {$t < T_n-W$}
				\STATE Exit the iteration and output $\hat{\mathbf{u}}_t$, if it is decoded successful or $l=I_{\max}$. 
				\ELSE
				\STATE Exit the iteration and output $\{\hat{\mathbf{u}}_t,\hat{\mathbf{u}}_{t+1},\cdots,\hat{\mathbf{u}}_{T_n-1}\}$, if they are decoded successful or $l=I_{\max}$. 
				\ENDIF
				\ENDFOR
				\ENDFOR
			\end{algorithmic}
		\end{algorithm}
		\vspace*{-8mm}
	\end{table}	
	
	In this section, we introduce the proposed windowed decoding to improve the detection of all sub-blocks. This is accomplished by iteratively performed two types of recursions, namely \textit{forward recursion} and \textit{backward recursion}. Define a decoding window of size $W$ as $W-1$ consecutive decoding layers, where $T_{\max}+1 \le W\le T_n$. An example of $W=3$ is shown in Fig. \ref{fig:DBICM_normal_graph}. The forward recursion means that the extrinsic information of the decoded delayed sub-blocks is passed to improve the detection on the undelayed sub-blocks from time $t$ to $t+W-1$ while the update on the undelayed sub-blocks follows Eq. (\ref{eq:demo_with_ap}). The backward recursion is backwardly passing the extrinsic information of the decoded undelayed sub-blocks, $\mathbf{c}_{t-T_{\bar{i'}}}(\bar{i'})$, to improve the detection on the delayed sub-blocks $\mathbf{c}_{t-T_{i'}}(i')$. To be specific, at time $t+1$, we first obtain $P_{b}(c^j_{t+1}({\bar{i'}}))$ following Eq. (\ref{eq:prob}). Then, $P_{b}(c^j_{t+1}({\bar{i'}}))$ is used to update the LLR of $c_{t}^j(i')$
	
	\begin{align} \label{eq:demo_with_ap_back}
		&L[c_{t}^j(i')|y_{t+1}^j,L_e[c_{t+1}^j({\bar{i'}})]] = \notag\\[2mm]  &\ln\left(\dfrac{\sum\limits_{b=0}^{1}\sum\limits_{\{\hat{x}_{t+1}^j\in\chi|\hat{c}_{t}^j({i'})=0,\hat{c}_{t+1}^j({\bar{i'}}) = b\}}\hspace*{-12mm}e^{-\frac{\norm{{y}_{t+1}^j-{\hat{x}_{t+1}^j}}^2}{2\sigma^2}}P_{b}(c^j_{t+1}({\bar{i'}}))}{\sum\limits_{b=0}^{1}\sum\limits_{\{\hat{x}_{t+1}^j\in\chi|\hat{c}_{t}^j({i'})=1,\hat{c}_{t+1}^j({\bar{i'}})=b\}}\hspace*{-12mm}e^{-\frac{\norm{{y}_{t+1}^j-{\hat{x}_{t+1}^j}}^2}{2\sigma^2}}P_{b}(c^j_{t+1}({\bar{i'}}))}\right).
	\end{align}  
	
	Use $L^{(0)}[\cdot]$ to denote the initial LLRs obtained from demapping without a priori information following Eq. (\ref{eq:demo_without_ap}). Furthermore, we have $L_e[\mathbf{c}_{-1}(i')]=\infty$ and $L_e[\mathbf{c}_{T_n}(\bar{i'})]=\infty$ due to initialization and termination of the DBICM transmission. In addition, we use $\overrightarrow{L}^{(l)}[\cdot]$ and $\overleftarrow{L}^{(l)}[\cdot]$ to denote the LLRs obtained from forward recursions and backward recursions, respectively, in the $l$-th windowed iteration. Specifically, the windowed decoding for DBICM is given in Algorithm \ref{algorithm:algorithm 1}.
	
	The proposed windowed decoding algorithm for DBICM can be divided into three main parts, namely the initial, the forward recursion, and the backward recursion. In the initial, which corresponds to Steps 2-8 in the algorithm, the initial LLR of all sub-blocks are detected. Steps 10-21 corresponds to the forward recursion, where soft information of the delayed sub-blocks forwardly propagate from nodes $\boxed{\text{M}} \rightarrow \boxed{\text{S/P}} \rightarrow \boxed{\Pi} \rightarrow \boxed{\text{C}} \rightarrow \boxed{\Pi} \rightarrow \boxed{\text{S/P}}$ at time $t$ to node $\boxed{\text{M}}$ at time $t+1$ to update the LLRs of the undelayed sub-blocks. Once the forward recursion finishes, backward recursion, corresponding to Steps 22-25, propagates the soft information of the undelayed sub-blocks from node $\boxed{\text{M}}$ at time $t$ back to nodes $\boxed{\text{S/P}} \rightarrow \boxed{\Pi} \rightarrow \boxed{\text{C}} \rightarrow \boxed{\Pi} \rightarrow \boxed{\text{S/P}} \rightarrow \boxed{\text{M}}$ at time $t-1$ to update the LLR of the delayed sub-blocks.
		\vspace*{-3mm}
	\subsection{Comparison}\label{subsec:comp}
	In this section, we compare our decoding with the original decoding in \cite{7593082} and the decoding algorithm for DBICM-ID \cite{8437452}. The original decoding of DBICM uses the extrinsic information from the delayed sub-blocks to improve the undelayed sub-blocks only once. In contrast, the proposed windowed decoding algorithm, with window size $W$, forwardly and backwardly detects $W$ received signals and decodes $W-1$ codewords in each iteration, such that the soft information is exchanged  between the delayed and undelayed sub-blocks to improve the detection on all sub-blocks. Thus, this provides performance improvements over the original decoding of DBICM. 
	
	Consider a decoding window $W$, the computational effort in computing the LLRs from demapping $W$ $2^m$-ary signal sequences is proportional to the constellation size and detection times. DBICM with the windowed decoding algorithm first computes the LLRs for $m$ bit-channels together in the initial detection. Then, in each iteration, the LLRs for the undelayed bit-channels and the delayed bit-channels are separately detected and updated in the forward and backward recursions. By considering the forward and backward recursions for $I_{\max}$ iterations and the operations needed for initialization, the proposed windowed decoding algorithm has a computational complexity of $\mathcal{O}\left((1+2I_{\max})W2^m\right)$. Consider DBICM-ID that has the same decoding latency as the windowed decoding of DBICM. Use $I'_{\max}$ to denote the maximum number of the iterations of detection and decoding in DBICM-ID. In each iteration, DBICM-ID distinctively updates the LLRs of each sub-block taking the extrinsic information from all other sub-blocks as a priori information. Therefore, the LLRs for $m$ bit-channels are separately computed via $m$ copies of detecting $2^m$ constellation points per iteration. In total, $\mathcal{O}(mWI^\prime_{\max}2^m)$ operations are required. For $I_{\max} = I^\prime_{\max}$, one can see that the detection complexity of DBICM-ID is much higher than the proposed windowed decoding, especially for large $m$. 
	
	%	We emphasize that the proposed windowed decoding algorithm for DBICM is different from the decoding algorithm proposed in \cite{8437452} for DBICM-ID. Similar to BICM-ID, DBICM-ID updates the LLRs of each sub-block by taking the extrinsic information from all other sub-blocks as the a priori information. Consider DBICM-ID that has the same decoding latency as the windowed decoding of DBICM. Regardless of the detection and decoding order, DBICM-ID iteratively detects $W$ received signal sequences and decodes $W-1$ codewords. We use $I'_{\max}$ to denote the maximum number of the iterations of detection and decoding in DBICM-ID. In each iteration, DBICM-ID distinctively updates the LLRs of each sub-block taking the extrinsic information from all other sub-blocks as a priori information. Therefore, DBICM-ID requires to detect each received signal sequence by $mI'_{\text{max}}$ times. On the other hand, the proposed windowed decoding updates the LLRs of all the undelayed/delayed sub-blocks in the forward/backward recursion. Therefore, the proposed windowed decoding algorithm only needs to detect each received signal sequence by $2I_{\text{max}}+1$ times. One can see that the detection complexity of DBICM-ID is much higher than the proposed windowed decoding, especially for large $m$.
	\vspace*{-2mm}
	\section{Numerical results} \label{sec:results}
	
	In this section, we show the simulation results of DBICM with windowed decoding on the AWGN channel for both the designed LDPC codes in \cite{9373632} and the off-the-shelf LDPC code. The performance is presented in terms of BER versus $E_b/N_0$. Here, $E_b$ stands for the average energy per source bit, which considers DBICM's spectral efficiency. To minimize the DBICM spectral efficiency loss, we consider a transmission frame\footnote{In the simulation, we require to receive at least $1000$ error frames for both BICM and DBICM at each $E_b/N_0$. For DBICM, this means that $T_n$ is at least $1001$.} $T_n \ge 1001$. For Gray labeled uniform $16$-QAM and $64$-QAM, delay schemes $\mathbf{T}=[0,1,0,1]$ and $\mathbf{T}=[0,0,1,0,0,1]$ are used, respectively. We include the BER performance of BICM, conventional DBICM decoding \cite{9373632}, and DBICM-ID for comparison. DBICM lower bound, which is based on the ideal assumption of the a priori information used for the detection of all sub-blocks in $\mathbf{x}_t$ being always correct, is provided as well. The designed LDPC codes with lengths $N=8,100$ and $N=12,000$ are constructed by following the degree profiles and the bit mapping design for Gray labeled uniform 16-QAM and 64-QAM, as shown in Tables III and IV in \cite{9373632}, respectively. Furthermore, we use the constrained progressive edge growth (PEG)-like algorithm in \cite{9373632} to obtain the LDPC codes with girth $6$.

	\begin{figure}[h]
		\vspace*{-3mm}
		\centering
		\includegraphics[width=0.5\textwidth]{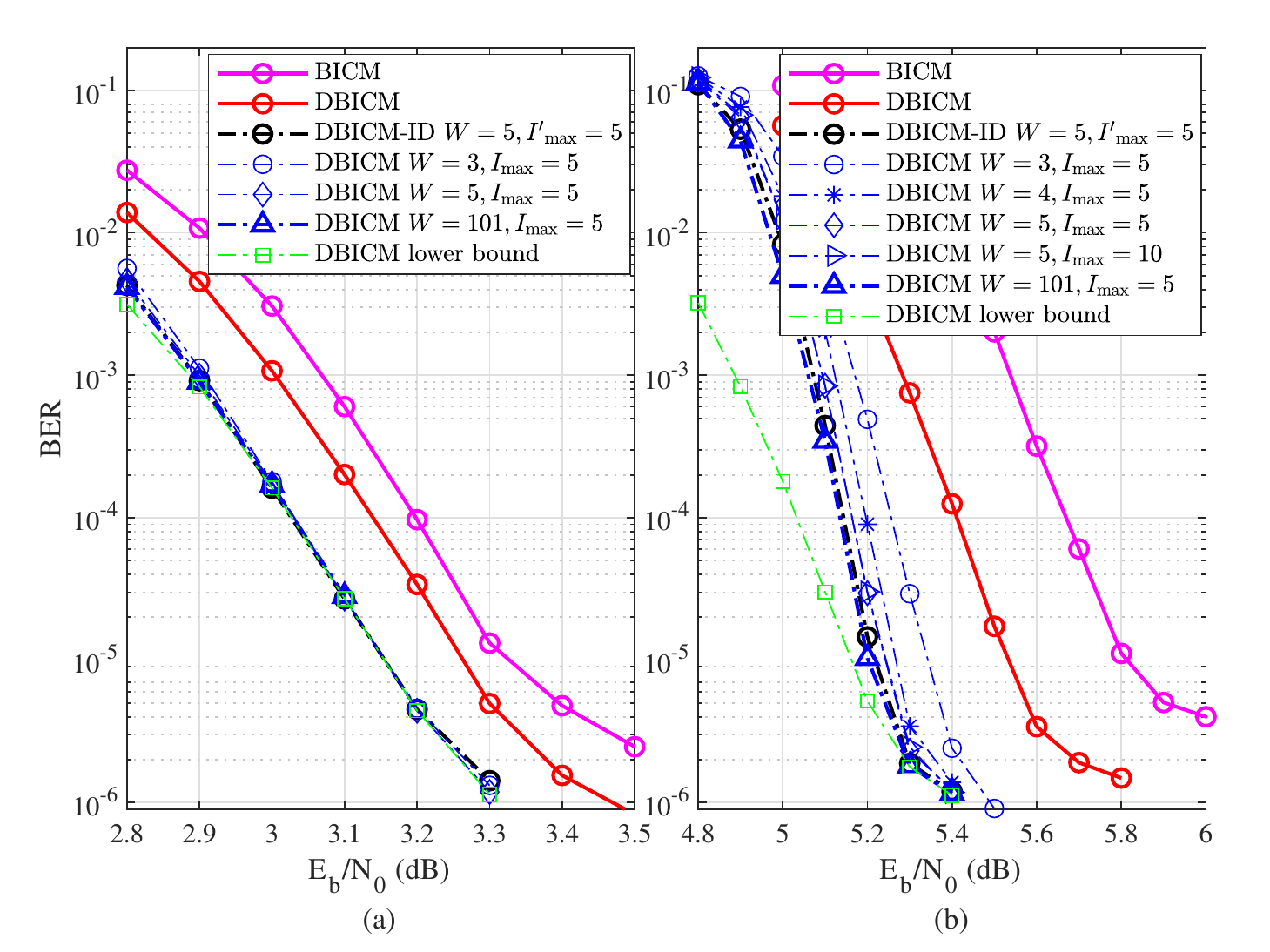}
		\vspace{-8mm}
		\caption{(a) BER of Gray labeled uniform 16-QAM using rate $1/2$ LDPC code designed in \cite{9373632} with various detection and decoding schemes, (b) BER of Gray labeled uniform 64-QAM using rate $1/2$ LDPC code designed in \cite{9373632} with various detection and decoding schemes.~~~~~~}
		\vspace*{-3mm}
		\label{fig:16_64qam_result}
	\end{figure}
	
	Fig. \ref{fig:16_64qam_result}(a) shows that windowed decoding for $16$-QAM DBICM, with $W=3$ and $I_{\text{max}}=5$, is sufficient to achieve good BER performance as its performance is very close to the DBICM lower bound. Also, DBICM with windowed decoding has a similar BER performance as DBICM-ID, but exhibit a lower detection complexity as described in Section \ref{subsec:comp}. The improvements from using windowed decoding over the conventional decoding for DBICM becomes large when $m$ is increased. As shown in Fig. \ref{fig:16_64qam_result}(b), for $64$-QAM, DBICM with windowed decoding outperforms its original decoding and BICM by around $0.30$ dB and $0.57$ dB, respectively at a BER of $10^{-5}$. Furthermore, the BER performance of windowed decoding with $W=5$ and $I_{\text{max}}=5$ approaches that of DBICM-ID with the same window size and maximum iteration number for all the considered $E_b/N_0$ and it is close to the DBICM lower bound in high $E_b/N_0$ region. In addition, it can also be noticed in Fig. \ref{fig:16_64qam_result}(b) that further increasing window size to $W=101$ or windowed iteration to $I_{\text{max}}=10$ has negligible performance improvement.

	\begin{figure}[h]
		\vspace*{-2mm}
		\centering
		\includegraphics[width=0.5\textwidth]{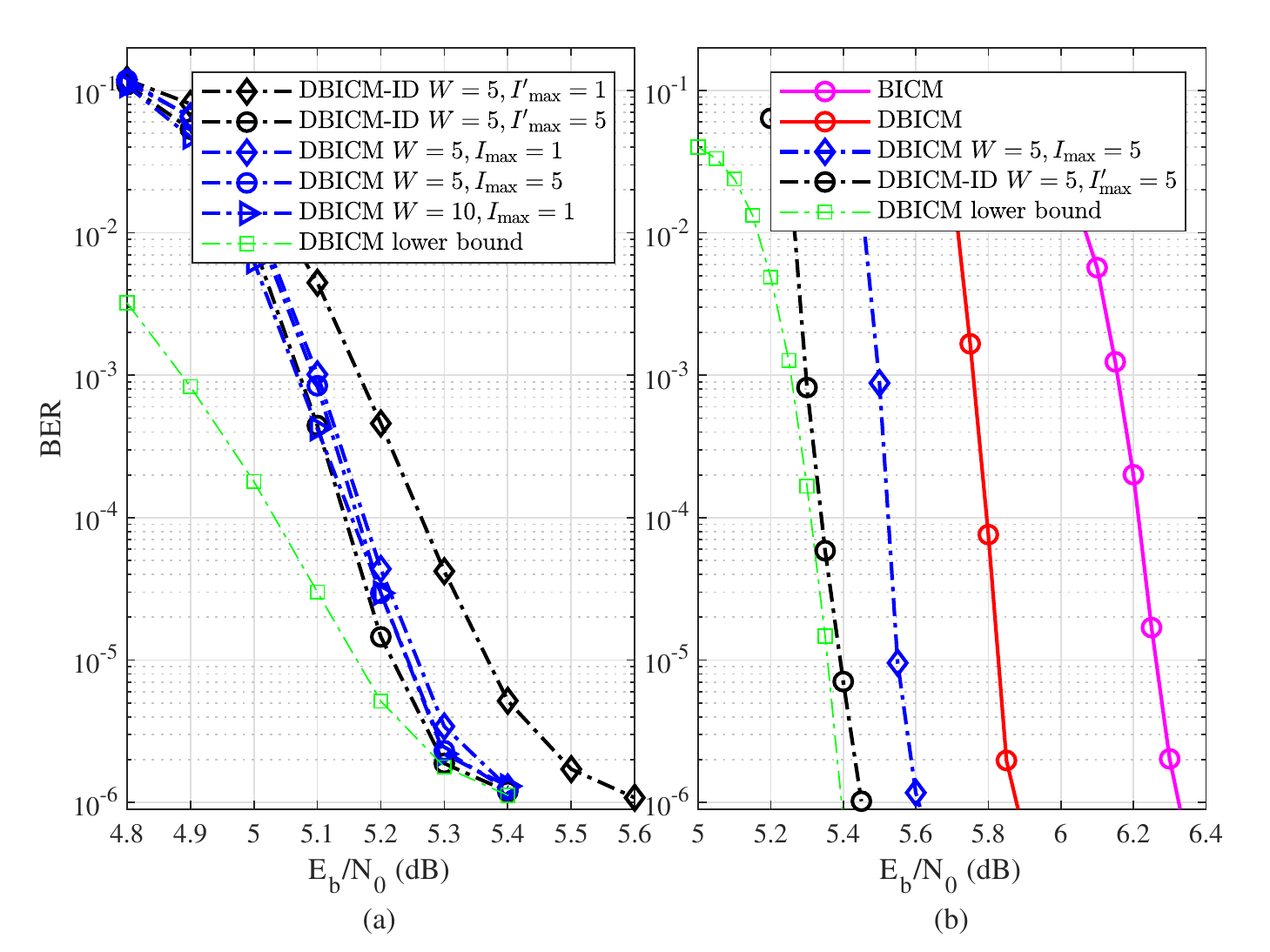}
		\vspace*{-8mm}
		\caption{(a) BER of rate $1/2$ LDPC coded uniform Gray labeled 64-QAM DBICM-ID and DBICM with windowed decoding under a variety of $I'_{\text{max}}$ and $I_{\text{max}}$, (b) BER of DVB-S2 32-APSK using rate $2/3$ DVB-S2 LDPC with various detection and decoding schemes.~~~~~~}
		\vspace*{-5mm}
		\label{fig:32apsk_64qam_result}
	\end{figure}
%	\vspace*{-10mm}
	
	Fig. \ref{fig:32apsk_64qam_result}(a) shows the comparison between windowed decoding for DBICM and DBICM-ID for a variety of window sizes and iterations. It is interesting to see that the performance of DBICM with windowed decoding at $I_{\text{max}}=1$ and $W=5$ is within $0.04$ dB to that of DBICM-ID with $I'_{\text{max}}=5$, $W=5$, and outperforms that of DBICM-ID with $I'_{\text{max}}=1$, $W=5$ by $0.11$ dB, at a BER of $10^{-5}$. It can be noticed from Fig. \ref{fig:32apsk_64qam_result}(a) that at $I_{\max}=1$, where computation resources are deficient, increasing window size $W$ shows limited BER performance improvement. In addition, compare Fig. \ref{fig:32apsk_64qam_result}(a) to Fig. \ref{fig:16_64qam_result}(b), DBICM with windowed decoding with $I_{\max}=1$ and $W=5$ outperforms DBICM-ID with $I^\prime_{\max}=1$ and $W=5$, the original decoding of DBICM, and BICM by $0.11$ dB, $0.23$ dB, and $0.53$ dB, respectively, at a BER of $10^{-5}$.   
	
	To show that the proposed windowed decoding also works well by using off-the-shelf LDPC codes, we adopt the standard modulations and LDPC codes from digital video broadcasting satellite second generation (DVB-S2) standard \cite{1566630}. Specifically, the LDPC code with rate $2/3$, length $64,800$, together with 32-ary amplitude and phase shift keying (APSK) modulation are used with random interleavers for both BICM and DBICM. For DBICM, the delay scheme is $\mathbf{T}=[0,0,1,0,1]$. As shown in Fig. \ref{fig:32apsk_64qam_result}(b), with $W=5$ and $I_{\max}=5$, the proposed windowed decoding algorithm exhibits performance improvements of about $0.69$ dB and $0.28$ dB over BICM and DBICM with the conventional decoding algorithm, respectively.
	
	In summary, using both the existing and designed LDPC codes with different codeword lengths, the proposed windowed decoding exhibits performance improvement over conventional DBICM decoding for various code rates. Furthermore, windowed decoding with a small $I_{\max}$ allows DBICM to approach the BER performance of DBICM-ID with a larger $I'_{\max}$. In addition, we show that the proposed windowed decoding provides a trade-off between the conventional decoding and DBICM-ID in terms of performance, complexity and latency.
%		\vspace*{-1mm}
	\section{Conclusion} \label{sec:con}
%		\vspace{-1mm}
	In this paper, we have proposed windowed decoding for DBICM in order to improve the detection of all sub-blocks. Specifically, the proposed windowed decoding makes use of the extrinsic information of the delayed and undelayed sub-blocks to aid the detection of undelayed and delayed sub-blocks, respectively. We show that the proposed decoding with small iterations in DBICM offers substantial gains over BICM and DBICM with its original decoding and has comparable performance to that of DBICM-ID with large iterations.
	
	\ifCLASSOPTIONcaptionsoff
	\newpage
	\fi
%			\vspace{-3mm}
	%	\linespread{1.5}\selectfont
	\bibliographystyle{ieeetr}
	\bibliography{pubs}

\end{document}